\documentstyle[aps,prl]{revtex}
\begin{document}
\thispagestyle{empty}
\pagestyle{plain}
\draft
\title{Anisotropic superexchange of a 90$^{\circ}$ Cu-O-Cu bond} 
\author{V. Yu. Yushankhai}
\address{Joint Institute for Nuclear Research, 141980 Dubna, Moscow 
Region,
Russia}
\author{R. Hayn}
\address{Institute for Theoretical Physics, TU Dresden, D-01062
Dresden, Germany}

\date{\today}

\maketitle

\begin{abstract}
The magnetic anisotropy af a rectangular Cu-O-Cu bond is investigated in
second order of the spin-orbit interaction. Such a bond is characteristic for
cuprates having edge sharing CuO$_2$ chains, and exists also in the
Cu$_3$O$_4$ plane or in ladder compounds. For a ferromagnetic coupling between
the copper spins an easy axis is found perpendicular to the copper oxygen
plaquettes in agreement with the experimental spin structure of
Li$_2$CuO$_2$. In addition, a pseudo-dipolar interaction is derived. Its
estimation in the case of the Cu$_3$O$_4$ plane (which is present for instance
in  
Ba$_2$Cu$_3$O$_4$Cl$_2$ or Sr$_2$Cu$_3$O$_4$Cl$_2$) gives a value 
which is however two orders of magnitude smaller than
the usual dipole-dipole interaction. 
\end{abstract}

\vspace*{3cm}

\noindent
\hspace*{2cm}PACS. 75.30.Gw - Magnetic anisotropy. \\
\hspace*{2cm}PACS. 74.72.Jt - Other cuprates. \\
\hspace*{2cm}PACS. 71.70.Ej - Spin-orbit coupling, Zeeman, Stark, and strain
splitting. \\ 

\newpage

To explain the rich magnetic behaviour of cuprates it is important to analyse
not only the dominant isotropic superexchange interaction but also the smaller
anisotropic terms due to the spin-orbit coupling \cite{Aharony}. In the case
of a 180$^{\circ}$ Cu-O-Cu bond, this explains for instance the 
``easy plane'' magnetism observed in the standard CuO$_2$ plane of orthorhombic
\cite{Kuzmin} 
or tetragonal \cite{Aharony2} cuprates. However, there is another important
structural element in many 
cuprates, namely a rectangular Cu-O-Cu bond. It is present in such cuprates
having a Cu$_3$O$_4$ plane (as in Ba$_2$Cu$_3$O$_4$Cl$_2$ \cite{Noro} or in
Sr$_2$Cu$_3$O$_4$Cl$_2$ \cite{Chu}), having 
CuO$_2$ chains (Li$_2$CuO$_2$ \cite{Sapina,Boehm}, CuGeO$_3$\cite{Geertsma})
or in 
ladder compounds \cite{Rice}. 
We are going to 
calculate here the parameters of the anisotropic superexchange interaction
for the ideal 90$^{\circ}$ bond. To incorporate lattice effects  
we investigate a Cu$_2$O$_6$ cluster (Fig.\ 1) - the common structural 
fragment of the cuprates mentioned. We choose as a representative example 
the Ba$_2$Cu$_3$O$_4$Cl$_2$ compound for which all important underlying  
parameters are derived from the extended multi-orbital tight-binding (TB)
analysis 
of its bandstructure. Some results of the TB analysis were presented in a 
previous publication \cite{Rosner}.
 
In the Cu$_3$O$_4$ plane of A$_2$Cu$_3$O$_4$Cl$_2$ (A=Ba,Sr) one has to
distinguish 
Cu$_A$ with a local orthorhombic symmetry and 
Cu$_B$ with local tetragonal symmetry. Lower than tetragonal local symmetry 
is obvious at the copper sites in the chains of Li$_2$CuO$_2$. This produces an 
additional feature - a weak higher order pseudodipole part of the magnetic
anisotropy  in these systems. In the leading orders of the perturbation 
theory the derived form of the anisotropic superexchange is similar for
A$_2$Cu$_3$O$_4$Cl$_2$ and Li$_2$CuO$_2$. 
For CuGeO$_3$ the analysis should be completed by taking into account the
influence of Ge side-groups \cite{Geertsma}.

The magnetic anisotropy can be calculated in
perturbation theory. In second order of the spin-orbit coupling $ \lambda (
\vec{L}_A \vec{S}_A 
+ \vec{L}_B \vec{S}_B ) $ with a characteristic parameter $\lambda=0.1$eV for
cuprates we obtain (see \cite{Yosida}):
\begin{equation}
\hat H^{\mbox{{\footnotesize aniso}}}_{AB}=\sum_{\mu\nu}
\left[ \Gamma^{\mu\nu} + \Gamma^{\nu\mu}
-\delta_{\mu\nu} (\sum_{\xi} \Gamma^{\xi\xi}) \right]
S_{A}^{\mu} S_{B}^{\nu} \; , 
\label{1}
\end{equation}
with
\begin{equation}
\Gamma^{\mu\nu}
= (\frac{\lambda}{2})^2
\sum_{mn}
\left\{ \frac{J(m_A,0_B;n_A,0_B)}
{(E_m^A - E_0^A)(E_n^A - E_0^A)}
+ ( A \leftrightarrow B ) \right\} L_{0m}^{\mu} L_{n0}^{\nu} \; , 
\label{2}
\end{equation}
where $J(m_A,0_B;n_A,0_B)$ is the integral of superexchange interaction 
between the groundstate
of Cu$_B$ and the excited states $m_A$ and $n_A$ of Cu$_A$. The non-zero
matrix elements of the orbital momentum are $L_{0x}^{x}=L_{0y}^{y}=i$,
$L_{0z}^{z}=-2i$ with the corresponding Hermitean conjugate ones. We will
denote the 3$d$ orbitals of Cu as follows: $|0\rangle=|x^2-y^2\rangle$,
$|1\rangle=|3z^2-r^2\rangle$,  $|z\rangle=|xy \rangle$, $|x\rangle=|yz
\rangle$ and $|y\rangle=|xz \rangle$ (in the x-y coordinate system of Fig.\
1). 
The small difference between the  Cu$_A$ and Cu$_B$ crystal field splitting,
which is 
inherent to the Cu$_3$O$_4$ plane, has no meaning for the present purposes and 
we assume with high accuracy $\Delta_n=E_n^A-E_0^A \simeq E_n^B-E_0^B$ and
$E_0^A \simeq E_0^B=E_0$. The values of $\Delta_n$ derived from the
bandstructure 
data \cite{Rosner} of Ba$_2$Cu$_3$O$_4$Cl$_2$ are listed in Table 1 and these
estimates are also relevant for other cuprates.

To analyse (\ref{1},\ref{2}) one has to calculate the different exchange 
integrals treating the Cu-O hopping  
($\sim t_{pd}$) and the O-O hopping
($\sim t_{pp}, t_{\pi\pi}, t_{p\pi}, t_{p_z}$) as perturbations. This is 
justified by the small ratios of these transfer integrals to the 
characteristic charge transfer energies $(E_{p,\pi,p_z}-E_0 )$. 
Together with the crystal field splitting the on-site interactions 
(involving direct and exchange Coulomb terms) at  copper and oxygen ions
are incorporated into the zero-order Hamiltonian. To avaluate the copper on-site
Coulomb interactions the Racah parameters are taken as in Table 1. For B and C
parameters the unscreened atomic values are used as explained elsewhere
\cite{Eskes}. The parameter A is adopted in such a way that the Coulomb
correlation $U_{00}^d=A+4B+3C$ for the groundstate Cu-orbital coincides with
the standard value (10.5 eV) given in Ref.\ \cite{Hybertsen}. Then one obtains
also the other direct ($U_{01}^d=A-4B+C$, $U_{0x,y}^d=A-2B+C$) and
exchange ($J_{01}^d=4B+C$, $J_{0x,y}^d=3B+C$) Coulomb integrals. For the 
correlation parameter $U_{p}$ at oxygen we take the standard value from 
Ref.\ \cite{Hybertsen} and $J_p$ from \cite{Geertsma}, which defines also 
$U_{p\pi}=U_p-2J_p$.
The values of the transfer integrals are derived from the tight-binding
analysis of the bandstructure \cite{Rosner}. Finally, for proper estimates 
of the
superexchange integrals one needs also the ``bare'' values for the charge
transfer energies $E_{p,\pi,p_z}-E_0$, instead of screened ones inferred from 
the bandstructure calculations. To solve this problem we have used a cluster
procedure \cite{Siura} connecting self-consistently the mean-field (screened)
values with ``bare'' ones for the on-site orbital energies.

 To perform calculations in a more effective way we introduce instead of the
original  in-plane oxygen $|p_{1,2} \rangle$- and $|\pi_{1,2}
\rangle$-orbitals their  
combinations, like 
\begin{equation}
\left.
\begin{array}{l}
|p\rangle \\
|\tilde p \rangle
\end{array} \right\}
= \frac{1}{\sqrt{2}}
\left[ |p_2\rangle \mp |p_1\rangle \right] \; , \qquad
\left.
\begin{array}{l}
|\pi\rangle \\
|\tilde \pi \rangle
\end{array} \right\}
= \frac{1}{\sqrt{2}}
\left[ |\pi_2\rangle \mp |\pi_1\rangle \right] \; ,
\end{equation}
with energies 
$\Delta_{p/\tilde p}=E_p-E_0\mp t_{pp}$ and
$\Delta_{\pi/\tilde \pi}=E_{\pi}-E_0\mp t_{\pi\pi}$, respectively (Table 1). 
For brevity we present below only the most important hybridization
interaction giving rise to the lowest order contributions (4th order in
$t_{pd}$) to the superexchange
\begin{eqnarray}
H_t^{(pd)} &=& t_{pd}
\left[ \sqrt{2} (d_{0A}^{\dagger} p + d_{0B}^{\dagger} \pi ) +
\sqrt{\frac{2}{3}} (d_{1A}^{\dagger} \tilde p - d_{1B}^{\dagger} \tilde \pi ) +
\sqrt{\frac{2}{3}} (d_{zA}^{\dagger} \tilde \pi - d_{zB}^{\dagger} \tilde p ) 
\right.
\nonumber \\
& & \left. + \sqrt{\frac{1}{3}} (d_{xA}^{\dagger} - d_{yB}^{\dagger}) p_{2z} +
\sqrt{\frac{1}{3}} (d_{yA}^{\dagger} + d_{xB}^{\dagger}) p_{1z} \right]
+ h.c. \; \quad .
\end{eqnarray}
Contributions to the superexchange due to the oxygen orbitals at the
right and the left cluster edges are mediated by O-O hybridization, which are
taken into account below as the 5-th and 6-th order corrections.

After these preliminaries we are able to estimate the different integrals
$J(m,0;n,0)$ entering into (\ref{2}). Let us start, however, with the
discussion of the basic isotropic superexchange between Cu$_A$ and Cu$_B$ in
the  groundstate. In our notation  that is given by 
$J(0_A,0_B;0_A,0_B)\equiv J_{AB}$. This exchange of 4th order in $t_{pd}$ 
and with a dominant ferromagnetic contribution due to the Hund's rule 
coupling $J_p$ at oxygen was estimated already several times (see 
\cite{Geertsma,Rice,Rosner,Siura}). In the next order (6th) of 
perturbation theory we find an antiferromagnetic contribution which arises 
due to the hopping $t_{p\pi}$ \cite{Rosner}. With the parameters chosen 
that sums up to a resulting ferromagnetic 
exchange of $J_{AB}$= -6 meV. 

Let us estimate now the other non-zero integrals responsible for the
anisotropy (\ref{1},\ref{2}). That has to be considered as  additional terms
on top  
of the dominant isotropic ferromagnetic term discussed above. First, we 
calculate 
$J(z_A,0_B;z_A,0_B)=J_{A,zz}^{(d)}+J_{A,zz}^{(p)}$ with two contributions 
in accordance with two possible intermediate virtual states in the 4th order
processes. These processes with the corresponding amplitudes and interaction 
parameters are shown schematically in Fig.\ 2. For the first contribution 
$\sim J_{A,zz}^{(p)}$, the singlet-triplet splitting in the virtual 
two-hole state,
with occupied oxygen $|\pi \rangle$- and $| \tilde \pi \rangle$-orbitals, is
due to the 
correlation interaction $U_p$. Then taking into account the other oxygen 
orbitals at the cluster edges, which gives the 5th order correction 
$\sim t_{pp}$, we obtain the result
\begin{equation}
J_{A,zz}^{(p)} = - \frac{4}{3} t_{pd}^4 
\left(\frac{1}{\Delta_{\pi}}+\frac{1}{\Delta_{\tilde \pi}} \right)^2
\left(\frac{1}{\Delta^T_{\pi\tilde\pi}}
-\frac{1}{\Delta^S_{\pi \tilde \pi}} \right)
\left( 1 + 2 t_{pp} ( \frac{1}{\Delta_p}-\frac{1}{\Delta_{\tilde p}} ) \right)
\; , 
\label{3}
\end{equation}
with  the triplet  and singlet energies  
$\Delta^T_{\pi \tilde \pi}=\Delta_{\pi}+\Delta_{\tilde\pi}$,
$\Delta^S_{\pi \tilde\pi}=\Delta_{\pi}+\Delta_{\tilde \pi} + U_p$, 
respectively. The 5th order correction gives only a 10 per cent decrease. In 
the intermediate state of the second contribution $J_{A,zz}^{(d)}$, the  
two holes  meet at  the same Cu$_B$  ion, but in different 
orbitals  $| 0_B \rangle $ and $| 1_B \rangle $. They interact due to the
Hund's coupling 
$\sim J_{01}^d$ to give
\begin{equation}
J_{A,zz}^{(d)}= 
- \frac{4}{9} \frac{t_{pd}^4}{\Delta_{\tilde\pi}^2}
\left[ \frac{1}{U_{01}^d-J_{01}^d} - \frac{1}{U_{01}^d+J_{01}^d} \right] 
\simeq
- \frac{4}{9} \frac{t_{pd}^4}{\Delta_{\tilde\pi}^2}
\frac{2 J_{01}^d}{(U_{01}^d)^2} \; , 
\label{4}
\end{equation}
without any correction of fifth order. The integral  $J(z_B,0_A;z_B,0_A)$ 
can be obtained by replacing $\pi,\tilde\pi\leftrightarrow p,\tilde p$. 
Analogously we find
$J(x_A,0_B;x_A,0_B)=J_{A,xx}^{(d)}+J_{A,xx}^{(p)}$ with 
\begin{equation}
J_{A,xx}^{(p)} \simeq - \frac{t_{pd}^4}{3}
\left( \frac{1}{\Delta_{p_z}} + \frac{1}{\varepsilon_{\pi}} \right)^2 
\frac{2J_p}{(\Delta_{p_z}+\varepsilon_{\pi}+U_{p\pi})^2}
\left( 1 + \frac{2t_{\pi\pi}}{\Delta_{p_z}} \right) \; , 
\label{5}
\end{equation}
\begin{equation}
J_{A,xx}^{(d)} \simeq - \frac{1}{9} \frac{t_{pd}^4}{\Delta_{p_z}^2}
\frac{2 J_{0y}^d}{(U_{0y}^d)^2} \; ,
\label{6}
\end{equation}
where $\Delta_{p_z}=E_{p_z}-E_{0}$ is the on-site energy of the out-of-plane
$|p_z \rangle$-orbital and 
$\varepsilon_{\pi}=\Delta_{\pi}+t_{\pi\pi}$. The same result also 
holds for $J(x_B,0_A;x_B,0_A)$ and for $J(y_A,0_B;y_A,0_B)=J(y_B,0_A;y_B,0_A)$
since $U_{0x}^d=U_{0y}^d$ 
and $J_{0x}^d=J_{0y}^d$. 

The pseudodipolar part of the superexchange for the Cu-O-Cu bond in a lattice 
arises due to deviations from the local tetragonal
symmetry.
Taking as a representative example the Cu$_3$O$_4$ plane this deviation occurs
at the Cu$_A$ sites.
The orthorhombic  energy splitting between the  
$|x^{\prime} \rangle$- and $|y^{\prime} \rangle$-orbitals 
(see the x$^{\prime}$-y$^{\prime}$ coordinate system of Fig.\ 1) 
is derived from the bandstructure results at the $\Gamma$ point to be 
$\delta \varepsilon_{xy} \approx 30$ meV. That corresponds to a mixed 
matrix element $\langle x | H | y\rangle = \delta \varepsilon_{xy}/2$ in 
the x-y coordinate system chosen in the
calculation. The second source for the pseudodipolar interaction is the
difference in the transfer integrals 
$\delta t_{p_z} = t_{p_z}^{\perp}-t_{p_z}^{\parallel}$,
between $|p_z \rangle$-orbitals perpendicular or parallel to the A-B bond
line. 
A rough estimate derived from  the bandstructure 
results \cite{Rosner} is found to be $\delta t_{p_z} \approx 40$ meV. 
Using these deviations from tetragonality as an
additional perturbation we obtain  the following estimate for the only  
non-zero off-diagonal exchange integrals 
\begin{equation}
J(x_A,0_B;y_A,0_B)=J(y_A,0_B;x_A,0_B) \sim \xi J^{(4)}(x_A,0_B;x_A,0_B)
\end{equation}
starting from the 4th order term of the diagonal exchange ((\ref{5}) and
(\ref{6}) without 5th order correction). The reduction factor is 
\begin{equation}
\label{8}
\xi = \frac{\delta \varepsilon_{xy}}{\Delta_x} + 2 \frac{\delta t_{p_z}}{\Delta
p_z} \; . 
\end{equation}
Using the parameters for Ba$_2$Cu$_3$O$_4$Cl$_2$ (see Table 1) 
we obtain finally  from (\ref{2})-(\ref{8}) the following numbers for the
anisotropic  
superexchange
\begin{equation}
\label{9}
\Gamma^{zz}=-0.65 \mbox{meV}, \quad
\Gamma^{xx}=\Gamma^{yy}=-10\mu \mbox{eV}, \quad
|\Gamma^{xy}| \sim 0.2 \mu \mbox{eV} \; . 
\end{equation}
In terms of the spin components parallel and perpendicular to the bond, the 
Hamiltonian (\ref{1}) takes the following form
\begin{equation}
\hat H^{\mbox{{\footnotesize aniso}}}_{AB}=
J_{\parallel} S_A^{\parallel} S_B^{\parallel} +
J_{\perp} S_A^{\perp} S_B^{\perp} +
J_z S_A^z S_B^z \; , 
\end{equation}
with $J_{\parallel}=-\Gamma^{zz} + 2 \Gamma^{xy}$, 
$J_{\perp}=-\Gamma^{zz} - 2 \Gamma^{xy}$ and 
$J_z=\Gamma^{zz} - 2 \Gamma^{xx}$. 
It should be noted that in spite of the nontetragonality of the particular 
bond under consideration we did not find in the leading orders of the 
perturbation procedure any process which could lead to  the 
Dzyaloshinskii-Moryia \cite{Yosida} spin coupling.

Let us discuss now the consequences for several cuprates and start with the
Cu$_3$O$_4$ plane.
First, one has to compare the derived value of the pseudodipolar part 
of the anisotropic superexchange 
$J_{pd}^{se}=(J_{\parallel}-J_{\perp})/2 = 2 \Gamma_{xy}$
with the estimate \cite{Chu} $J_{dd} \approx 20$ $\mu$eV for the usual 
dipole-dipole interaction for the Cu$_A$-Cu$_B$ bond. The small value of
$J_{pd}^{se}$ supports
the suggestion
\cite{Chu} that the weak ferromagnetic moment observable in 
A$_2$Cu$_3$O$_4$Cl$_2$ 
is mainly due to the usual dipole-dipole interaction. Effects of the other
kind of anisotropy given by $\Delta J = (J_{\parallel}+J_{\perp})/2 - J_z =
-2(\Gamma_{zz}- \Gamma_{xx})>0$ could be revealed only in a more general
context. Actually, to predict the low-temperature magnetic structure in
the above compound one has to take into account the main interations
(including the isotropic and anisotropic terms) for Cu$_A$-Cu$_A$ and
Cu$_B$-Cu$_B$ bonds in the Cu$_3$O$_4$ plane. This requires a more extended
analysis for these two interacting antiferromagnetic subsystems  
\cite{Aharony,Chu}.
   
At the same time, in Li$_2$CuO$_2$, the estimate $\Delta J >0$ is
important to explain the magnetic ordering. 
Based on preliminary  bandstructure results \cite{Rosner2} one may infer
that the relevant parameters for
Li$_2$CuO$_2$ will be only slightly changed in comparison to Table
1. Therefore, it can be expected that the above result for $\Delta J$ remains
qualitatively correct. At low temperatures, for a chain with ferromagnetically
aligned spins, the anisotropy $\Delta J >0$ prefers a spin direction parallel
to the z axis, i.e.\ perpendicular to the chain plaquettes. That agrees
with the experimental situation \cite{Sapina,Boehm}. The measured energy gap
in the spin wave spectrum \cite{Boehm} leads to a uniaxial anisotropy which is
in agreement with our estimate for $\Delta J$. But the values of the isotropic
superexchange coupling reported in \cite{Boehm} are about one order of
magnitude smaller than that estimated from the bandstructure calculation
\cite{Pickett} and characteristic also for other cuprates. It would be
interesting to reanalyse the spin wave spectrum starting from a spin
Hamiltonian based on the microscopic electronic structure. Some additional
complications should be noted as well. Actually, a magnetic frustration occurs
in Li$_2$CuO$_2$  since the isotropic antiferromagnetic
superexchange for second neighbours within a chain is rather strong. That is
expected 
\cite{Rosner2,Pickett} to exceed even the ferromagnetic coupling for nearest 
neighbour spins. The stability of the overall magnetic structure in
Li$_2$CuO$_2$ with antialigned ferromagnetic chains can be explained due
to  a considerable three dimensional spin coupling in this compound
\cite{Rosner2,Pickett}. 

We thank M.\ Kuzmin, J.\ Schulenburg, K.H.\ M\"uller, M.\ Wolf and P.\
Oppeneer for useful 
discussions. After completing our calculations we became aware that similar
results were obtained by S.\ Tornow, O.\ Entin-Wohlmann and A.\ Aharony using 
a different technique 
and neglecting the local orthorhombic distortion \cite{Tornow} and we thank
them also for discussion and for their interest in our work.

\newpage

\begin{center}
{\Large\bf TABLES}
\end{center}

TABLE 1: Parameters (in eV) to calculate the magnetic anisotropy.

\vspace*{1cm}

\begin{tabular}{||c|c|c||c|c|c||c|c|c||}
\hline
\hline
\multicolumn{3}{||c|}{copper crystal field splitting}&
\multicolumn{3}{|c|}{transfer integrals} &
\multicolumn{3}{|c||}{oxygen correlations}\\
\hline
$\Delta_1$ & $\Delta_z$ & $\Delta_{x,y}$ &
$t_{pd}$   & $t_{pp}$   & $t_{\pi \pi}$  &
$U_p$      & $U_{p\pi}$ & $J_p$          \\
0.65     & 1.08     &  1.32        &
1.40     & 0.80     &  0.45        &
4.0      & 3.2      &  0.4         \\
\hline
\hline
\end{tabular}

\vspace*{0.8cm}

\begin{tabular}{||c|c|c|c|c||c|c|c||}
\hline
\hline
\multicolumn{5}{||c|}{oxygen on-site energies}&
\multicolumn{3}{|c||}{copper correlations} \\
\hline
$\Delta_{\pi}$        & $\Delta_{p}$        & $\Delta_{p_z}$ & 
$\Delta_{\tilde \pi}$ & $\Delta_{\tilde p}$ &
$A$          & $B$        & $C$  \\
3.5  & 3.9  &  4.0  &  4.4  & 5.5  &
8.16 & 0.15  &  0.58 \\
\hline
\hline
\end{tabular}

\vspace*{2cm}

\begin{center}
{\Large\bf FIGURES}
\end{center}

\vspace*{1cm}

FIG.\ 1: The Cu$_2$O$_6$ cluster used to calculate the magnetic anisotropy
(full circles: copper, open circles: oxygen) with the ground state orbitals
$|0\rangle$ at copper and the in-plane oxygen orbitals (full lines: $|p
\rangle$-orbitals and broken lines $|\pi \rangle$-orbitals). Also shown are
the two 
coordinate 
systems mentioned in the text. 

\vspace*{0.7cm}

FIG.\ 2: Examples for 4th order processes which contribute to $J_{A,zz}^{(d)}$
and $J_{A,zz}^{(p)}$.

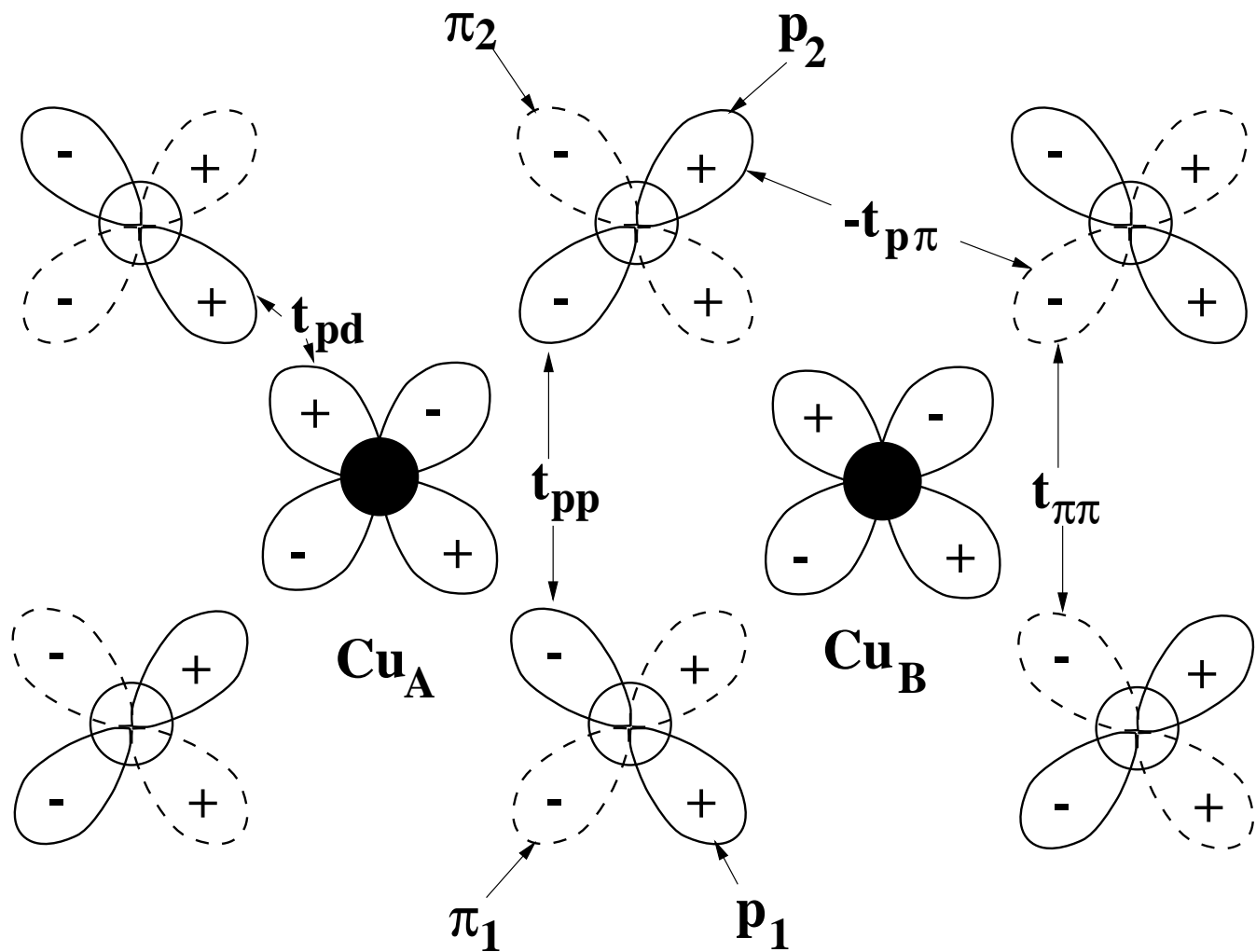
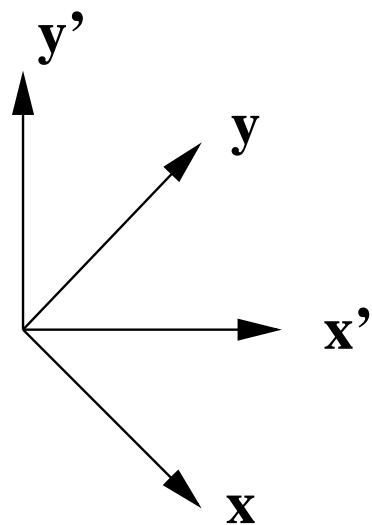
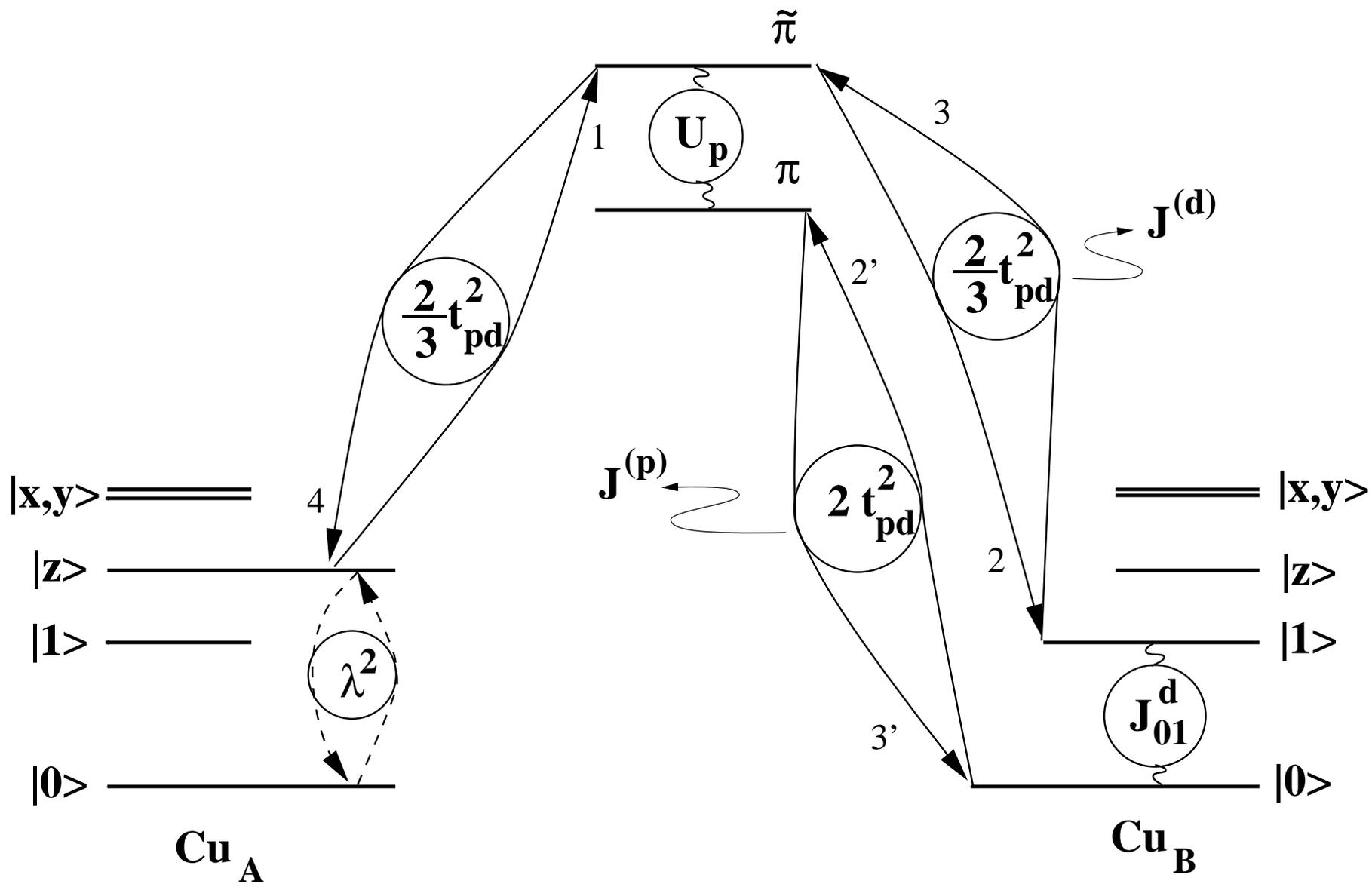
\end{document}